# Querying Spreadsheets: An Empirical Study

Jácome Cunha ∗†, João Paulo Fernandes ∗‡, Rui Pereira∗◊ , and João Saraiva ∗◊
∗ HASLab/INESC TEC ◊ Universidade do Minho, Portugal
† Universidade Nova de Lisboa, Portugal
‡ RELEASE, Universidade da Beira Interior, Portugal
jacome@fct.unl.pt, jpf@di.ubi.pt, {ruipereira,jas}@di.uminho.pt

*Abstract*— **One of the most important assets of any company is being able to easily access information on itself and on its business. In this line, it has been observed that this important information is often stored in one of the millions of spreadsheets created every year, due to simplicity in using and manipulating such an artifact.**
**Unfortunately, in many cases it is quite difficult to retrieve the intended information from a spreadsheet: information is often stored in a huge unstructured matrix, with no care for readability or comprehensiveness.**
**In an attempt to aid users in the task of extracting information from a spreadsheet, researchers have been working on models, languages and tools to query.**
**In this paper we present an empirical study evaluating such proposals assessing their usage to query spreadsheets. We investigate the use of the Google Query Function, textual model-driven querying, and visual model-driven querying. To compare these different querying approaches we present an empirical study whose results show that the end-users' productivity increases when using model-driven queries, specially using its visual representation.**

*Keywords- spreadsheets; querying; model-driven querying; textual querying; visual querying; empirical validation;*

## I. INTRODUCTION

Spreadsheet systems are one of the most widely used software systems. Spreadsheets were introduced as a simple, visual, human-friendly, and easy-to-use software system to allow non-professional programmers, usually called end users [13], to express mathematical equations.

Spreadsheets, however, rapidly evolved into large and complex systems used in many different contexts. Today, spreadsheets can be found nearly everywhere in companies and are used for a variety of purposes. In fact, spreadsheet systems are not only used to implement spreadsheet specific tasks (for example, to implement companies' budgets, to express risk assessments for insurance companies, etc.), but also to manipulate large and complex data: for example to collect and group information from different systems, to perform operations to enrich or simplify data, to present data in human-friendly form, or to transform data coming from one system to the format required by another.

Actually, spreadsheets can be seen as the "poor man's database", used to store data in a simple and visual way, but not having the specific features of a database system. It is surprising to see that 56% [14] of spreadsheets in the large EUSES corpus do not contain formulas, only data! In order to manipulate such data, spreadsheets need to provide the data normalization mechanisms and a powerful query language, as databases do.

## II. SPREADSHEET QUERYING

Some techniques have been made to query spreadsheets. In this paper, we will look at one technique developed by Google, the QUERY function [6], and two model-driven querying systems, namely, QuerySheet and Graphical-QuerySheet (textual and graphical based respectively).

Before presenting these techniques, let us introduce a spreadsheet to be used as a running example throughout this paper.

Fig. 1 presents part of a spreadsheet used to store information relative to the budget of a research group. This spreadsheet contains information about the *Category* of budget used (such as *Travel* or *Meals*) and the *Year*. The relationship between *Category* and *Year* provides us with the information on the *Quantity*, *Cost*, and the *Total Costs* (defined by spreadsheet formulas), per year per category.

|   | A | B | C | D | E | F | G | H | I |
|---|---|---|---|---|---|---|---|---|---|
| 1 | Budget | | Year | | | Year | | | Year |
| 2 | | | 2005 | | | 2006 | | | 200 |
| 3 | Category | Name | Qnty | Cost | Total | Qnty | Cost | Total | Qnty |
| 4 | | Travel | 2 | 525 | 1050 | 3 | 360 | 1080 | |
| 5 | | Accomodation | 4 | 120 | 480 | 9 | 115 | 1035 | |
| 6 | | Meals | 6 | 25 | 150 | 18 | 30 | 540 | |

Figure 1. Spreadsheet example.

Using the example spreadsheet, we will be answering the following simple question:

**Question**: In what category did we have the most expenses in the last 5 years?

In the following paragraphs we briefly introduce three different systems to query spreadsheets and answer this question.

### A. Google QUERY function

Google provides users a querying function called QUERY. This QUERY function performs a query, using a SQL-like syntax [5], over an array of values such as the Google Docs spreadsheets, in which the function is built in. Google's QUERY function (GQF) is a two argument function, consisting of a range and query string. The range argument is used to state the range of the data cells to be queried, for example A1:Q13 in our spreadsheet. The query string is the actual SQL-like query written by the user.

While the GQF is a powerful query function, it still has some flaws. To run this function, the user needs to represent his/her spreadsheet information in a single table, with each attribute represented in each column (in other words, with headers). This means that someone, who has their spreadsheet divided into various entities with or without relations, would first need to manually denormalize their data (as shown in Fig. 2). Obviously, such a process is difficult enough for someone experienced in data normalization/denormalization techniques [1], let alone for end users.

Figure 2. Spreadsheet example denormalized.

Along with the difficulty in managing the data in such a way, the GQF has another flaw. When writing the QUERY string, the user must write column letters in the query, instead of column names/labels as is normal in database querying. As one might assume, this can get confusing, counter-intuitive, unproductive, and almost impossible to understand what the query is supposed to do. Even with our small spreadsheet example, query construction would be impossible without looking at the denormalized data.

Moreover, this form of manually defining the spreadsheet range and query string with column letters brings about one final flaw: no evolution support. Since the queries do not adapt/evolve when the spreadsheet data evolves, simple tasks such as adding a new column (in other words a new attribute), or new rows of information, we may turn a query invalid or incorrect because the data changed positioning in the spreadsheet.

Regardless, the query engine is very efficient, being able to handle very big spreadsheets. So if we wanted to answer our previous question, we would have to write in a spreadsheet cell, the following query function:

=query(A1:E58;"SELECT B, sum(E) WHERE A >= 2010
       GROUP BY B ORDER BY sum(E) DESC
       LIMIT 1")

B. QuerySheet

To overcome the issues identified with the GQF, researchers turned to model-driven engineering methodologies [7-10] to design a query language and system for spreadsheets. In this case model-driven spreadsheet models were used, specifically ClassSheets [3,4,10]: a high-level and object-oriented formalism, using the notion of classes and attributes, to express business logic spreadsheet data. Using ClassSheets, one can define the business logic of spreadsheet data in a concise and abstract manner. This results in users being able to understand, evolve, and maintain complex spreadsheets by just analyzing the ClassSheet models, avoiding the need to look at large and complex data. Indeed, as shown in [12], users need a bridge between spreadsheet data and the real world.

To show a ClassSheet example, and the corresponding spreadsheet model for the model-driven querying systems, we present that specifies the Budget example shown in Fig. 1. In this ClassSheet model, a Budget has a *Category* (with a *Name* attribute) and *Year* class (with a *Year* attribute), expanding vertically and horizontally, respectively. The joining of these two gives us a *Quantity*, *Cost*, and *Total* of a *Category* in a given *Year*, each with their own default values. The corresponding spreadsheet instance conforms to the ClassSheet model as shown in Fig. 3.

Figure 3. Spreadsheet example.

Using this spreadsheet model concept, researchers were able to design a querying language based on the attributes/labels in classes, as done in the database realm when using attribute names from tables. This querying system was named QuerySheet, and is integrated in the MDSheet framework, a model-driven spreadsheet environment with all the mechanisms to handle models, instances, and evolution.

This system is based on using Google's QUERY function engine, automatically denormalizing the spreadsheet models, translating the model-driven query to the query function counterpart (using generative and transformational techniques [11]) and sending both translated function and denormalized data to be executed on Google's system. Afterwards, it applies model inference techniques to produce a model and corresponding conformed instance for the results, allowing composable queries.

The translation automatically calculates the range of the spreadsheet data and the appropriate column letters, thus allowing evolution. Querying denormalized data also brings in many known problems, due to the redundant data, which our system also automatically handles, using various mechanisms to ensure that querying the denormalized spreadsheet data does not lead to incorrect results.

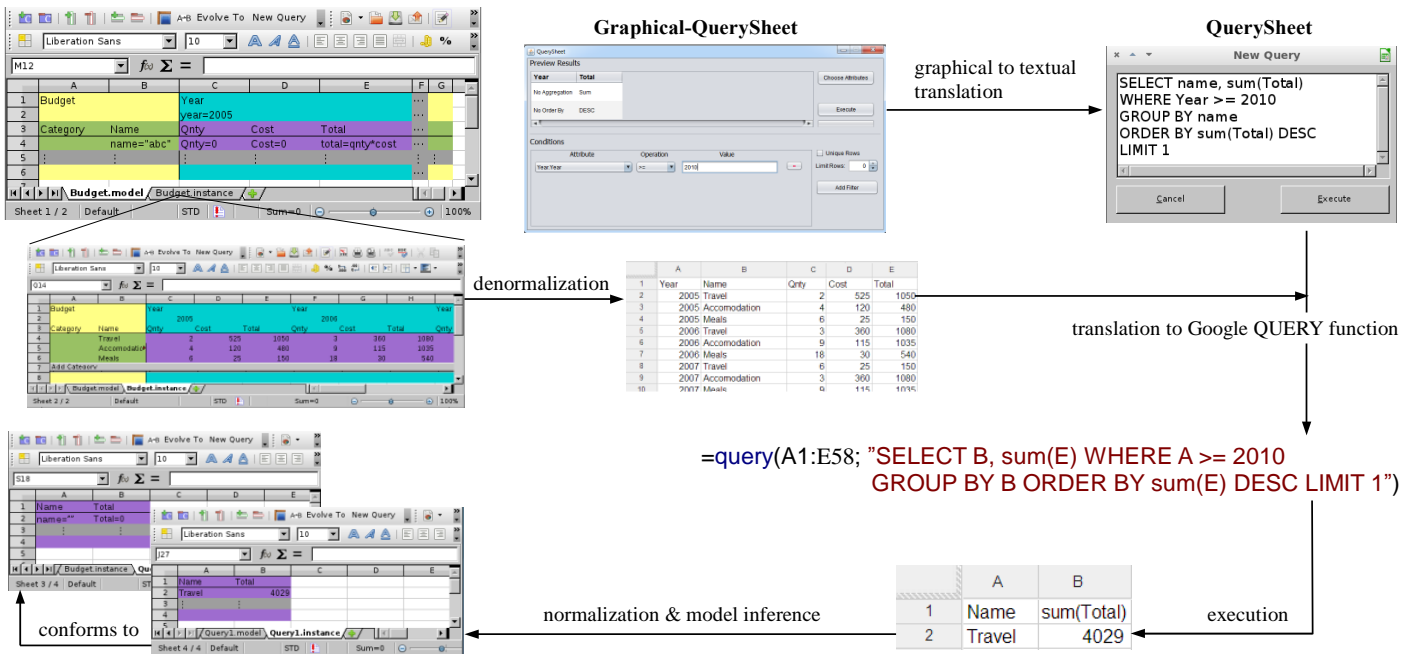

Figure 4. QuerySheet and Graphical-QuerySheet architecture

Therefore, answering the previous question would be as simple as looking only at the ClassSheet model, and writing the following query:

> SELECT name, sum(Total)
> WHERE Year >= 2010
> GROUP BY name
> ORDER BY sum(Total) DESC
> LIMIT 1

The architecture of this system can be seen in Fig. 4.

### C. Graphical-QuerySheet

Although QuerySheet overcomes all limitations of using GQF, namely data denormalization and column references, the fact is that for end users it is difficult to write SQL [5] sentences.

To overcome this issue, it has been developed a graphical user interface for visual query construction. This interface was aimed to make querying simpler for end users or users with less experience in writing SQL queries. The interface was made to be intuitive to use, both for experienced and non-experienced users. Its interface also intends to reduce the amount of errors (at least due to query syntax and attribute's names), since it lets the user to choose the attributes based on the spreadsheet's model, as shown in Fig. 5.

This interactive graphical query building interface, named Graphical-QuerySheet, was built on top of QuerySheet. When the user executes his/her query, the visual language is translated to our model-driven querying language presented in the previous section (and shown in. Fig. 4), and the remaining process maintains the same. The model-driven query is translated into Google's QUERY function, the data is denormalized, and both are executed on Google's system. An inferred model is produced from the resulting data, and made into an instance in conformance with the inferred model.

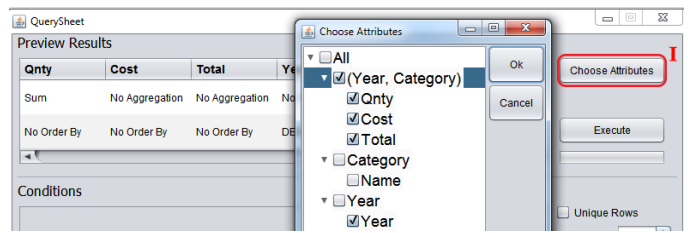

Figure 5. Graphical select.

Using this graphical querying interface, one can easily answer the question, as shown in Fig. 6, the following way:

1. Click on Choose Attributes and check Name and Total;
2. Click on the aggregation combo box (it becomes visible when using the tool) under Total and choose Sum;

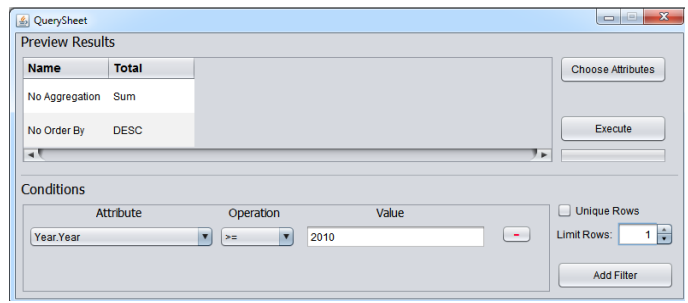

Figure 6. Graphical query.

3. Click on the order by combo box under Total and choose DESC;
4. Click on Add Filter;
5. Select the Year.Year attribute and greater or equal to operation using the combo boxes, and fill in 2010 in the text box;
6. Click Execute.

## III. EMPIRICAL STUDY

In order to assess and compare Google's QUERY function, QuerySheet, and Graphical-QuerySheet in practice, we planned and executed an empirical study with 14 users, obtaining and recording the results of their experiences, productivity, and feedback.

All participants were male, between ages 19-28, with some form of background in computer sciences/informatics. Their knowledge of SQL also varied between no/little knowledge to very experienced users. All have previously worked with spreadsheets, with different levels of experience.

These 14 participants were randomly divided into two groups. One group was to test GQF vs. QuerySheet, and the other to test GQF vs. Graphical-QuerySheet.

This empirical study was done with one participant at a time, in a think-aloud session. Doing this allowed us to see each participant using the systems, learn the difficulties they were facing, and know where the systems need to be improved.

For the study, we used a real-life spreadsheet obtained from, and with permission to use, the food bank in our hometown. This spreadsheet stored information regarding the distributions of basic products to specific institutions, containing information on 85 institutions, with 14 types of products, and over 1190 lines of unique information.

We denormalized the spreadsheet information for the participants to use with Google's QUERY function, and prepared the spreadsheet models and conformed instance in the MDSheet environment. Due to revealing private information in the spreadsheet, only the spreadsheet model, which is also the same one used in the study, is presented in Fig. 7.

As we can see in the model, and consequently in the actual spreadsheet, the *Distribution* class is composed of an *Institution* class and *Product* class. The *Institution* class has a *Code*, *Name*, *Lunch* (units used for lunch and snacks), and *Dinner* (units used for lunch and dinner) attribute. The *Product* class has a *Name*, *Code*, and *Stock* (representing the amount of a specific product in stock) attribute. The relationship between both of these two classes gives us the information on the quantity *Distributed* of a specific *Product* to a specific *Institution*.

After thoroughly explaining to the participants how the information was represented (in the denormalized and model-driven format), and how to correctly interpret the information, participants started answering the questions of the study.

*A. Execution*

In the study, we asked participants to implement queries to answer the following four questions regarding the information present in the distributions spreadsheet:

1. What is the total distributed for each product?
2. What is the total stock?
3. What are the names of each institution without repetitions?
4. Which were the products with more than 500 units distributed, and which institution were they delivered to?

In answering each question, the participants had to implement a query using both systems (either GQF vs. QuerySheet or GQF vs. Graphical-QuerySheet); alternating between the starting systems (the initial starting system was chosen by each participant). This alternation was introduced in the study so the potential learning from answering a question in one system would not interfere with the results from the second system. Since the order of the systems used to answer the questions alternated between each question, and for each participant, the potential learning can be ignored for both systems.

The users were asked to write down the starting time after carefully reading each question, and the time after the queries were executed with no errors (the correctness of the queries and results were analyzed afterwards). They would then re-read the question, and repeat the same process, writing down the starting and ending time. It is also to note, that the differences in the running performance of Google's QUERY function compared to the QuerySheet/Graphical-QuerySheet system are negligible as all are almost instantaneous).

After completing each question with both systems, they were asked to answer a short questionnaire to choose which system they felt was more:

- Intuitive
- Faster (to construct the queries)
- Easier (to construct the queries)
- Understandable (being able to easily interpret, and explain the constructed queries)

To conclude the study, the participants answered which system they preferred, why, and what advantages/disadvantages existed between the two compared systems. In this case they could write free text.

Figure 7. A model-driven spreadsheet representing food bank distributions.

## B. Results

The results we obtained from our study were gathered and analyzed, and are presented in this section.

Fig. 8 and Fig. 9 can be interpreted as follows: The Y-Axis represents the average number of minutes the participants took to answer the questions. The X-Axis represents the question the participants answered. The green[1] bars represent the Google QUERY function, and the blue bars represent the QuerySheet and Graphical-QuerySheet system respectively.

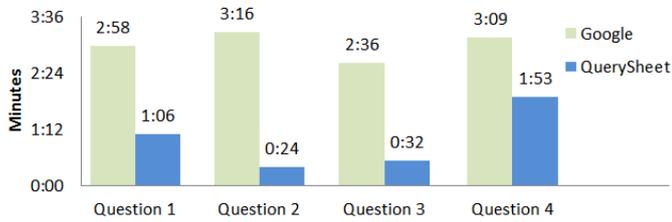

Figure 8. GQF v. QuerySheet results.

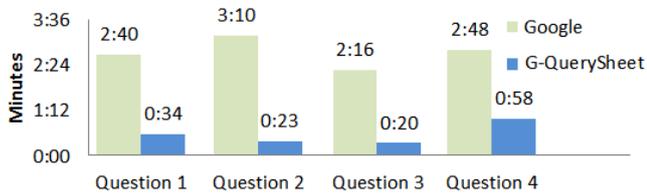

Figure 9. GQF v. Graphical-QuerySheet results.

For easy referencing, we will refer to the first study group, Google QUERY function vs. QuerySheet, as Group A, and for the second study group, Google QUERY function vs. Graphical-QuerySheet, as Group B.

As we can see in both groups, the participants spent substantially less time to construct the queries, and in turn were more productive, using the model-driven approach (both QuerySheet and Graphical-QuerySheet). In Group A, participants spent 62.8% to 87.8% less time, averaging to an overall of 67.5%. Looking at the graphical querying approach, in Group B, the participants spent 65.5% to 87.9% less time, averaging to an overall 79.4%.

Almost all chose the model-driven querying approaches, in terms of the four previously mentioned points (Intuitive, Faster, Easier, and Understandable). In Group A, 111 out of 112 (4 points * 4 questions * 7 participants) chose QuerySheet. While in Group B 104 out of 112 chose Graphical-QuerySheet, providing us with interesting information, allowing us to detect some of the drawbacks of the graphical system, which will be explained further on.

In general the query error rates were low. There were 9 (out of 11) and 6 (out of 7) errors, Group A and B respectively, using Google's QUERY function.

---

[1] We assume colors are visible in the digital version of this paper.

These errors varied between incorrect column letters chosen, bad query construction and incorrect ranges.

The 3 remaining errors occurred when using the model-driven queries were due to bad query construction. Those same errors were also repeated on their Google QUERY function counterpart. With this, we can conclude that these model-driven based errors, and in turn the Google QUERY function counterpart, were most likely due to inexperience in query construction.

Both verbal and textual comments by the participants were also very interesting. Both groups shared some common comments, in favor of the model-driven approaches such as (the citations are translated from Portuguese, the language in which the study was performed):

- *"Using attribute names instead of column letters is simple and natural"*
- *"Not having to calculate the range saves time and headaches"*
- *"Having a distinct clause is extremely useful and makes querying faster"*
- *"Very intuitive, easy to construct a query, and easy to understand"*

Nevertheless, there were some differences. For example, participants from Group A commented on how:

- *"the usage of models helped a lot in building queries"*
- *"it was much more intuitive to look at the model, and easily understand the layout and refer to the labels"*

While participants in Group B made no comment whatsoever. As a matter of fact, Group B participants never looked at the model, other than the first time when presented. We were able to quickly understand that the attribute selection window in Graphical-QuerySheet intuitively represented the spreadsheet models in a way that users only needed to look at that window. Indeed this is a very interesting lesson as it can be used in other tools. Regardless of this fact, we can still say that the spreadsheet models are essential in the graphical system.

There was another difference we noticed in the comments. The participants from Group B said:

- *"No need to know SQL, a normal user like myself can quickly and easily construct queries"*
- *"I do not need to worry about using group by when I aggregate, Graphical-QuerySheet does it automatically for me"*

We can say that, disregarding the fact of having an automatic *Group By* (which obviously makes query construction simpler), that Graphical-QuerySheet's interface abstracts the SQL language in a simple visual language. This allows users to construct queries without needing any knowledge of the syntax, or anything "behind the scenes", making it very human-friendly and end-user focused.

Finally, all participants chose the model-driven approaches as their preferred system.

## C. Drawbacks

Even though both model-driven approaches produced very good results, we did learn what aspects are lacking in the two querying systems, along with other interesting points which differed between the more experienced SQL users and those with only basic or no knowledge.

For example, in Group A, we often realized that, independent of which system was being used, some participants would have difficulty in writing the textual based queries. This was recurring among the participants with less experience in SQL, and in some cases, even among the more experienced ones. This was due to having to remember the written syntax.

Unlike the previous example though, the more experienced users in Group B found it rather awkward (and during the study took a while) to have to choose an attribute to be able to use the *Order By* option, while the less experienced users quickly and intuitively chose the attributes to use the clause. We learned that this was due to the less experienced users being able to abstract themselves from the typical textual SQL language, while the more experienced ones had a deeper connection, and in turn viewed the attribute preview panel as the *Select* clause. While this may be viewed as a drawback from a certain point-of-view, since the goal graphical system wanted to simplify query construction for an end user, we believe this to actually be positive.

As previously mentioned, there were few cases in Group B where the participants preferred Google's QUERY function (6 out of 112). All of these cases were regarding which system was more understandable. The participants who chose Google's QUERY function stated that they were more used to reading a line of text, instead of looking around an interface to interpret the information. This reason is mostly out of habit (since the participants were all from informatics/computer sciences). Nevertheless, the graphical interface can be improved to be more legible, or even provide a quick textual preview of the query.

There were two cases where the participants preferred neither in regards to which was Easier, both during Question 4 (where many clauses were needed such as *Order By*, *Limit*, *Where*). In these cases, they believed it was equally difficult to construct in either system due to having to use many conditions.

## IV. CONCLUSION

In this paper we have presented an empirical study and evaluation on three querying systems, two of which are model-driven based. Overall, we believe we can say that the model-driven querying approaches have proven themselves to be much more efficient and easier to use than their counter part modeless.

We have also shown that users with less experience in SQL not only have an easier time with, but also prefer, a graphical based querying system. At the same time, some users with more experience in SQL might find it easier to interpret the textual representation, out of habit. Nevertheless, they do admit the graphical model-driven approach is less complicated and overall the easier way to construct spreadsheet queries.

Indeed it would be quite interesting to perform a similar study for the database realm. Given the similarities between the querying systems in both worlds, very similar study settings should apply.

## REFERENCES

[1] Maier, D.: The Theory of Relational Databases. Computer Science Press (1983)

[2] Hainaut, J.L.: The transformational approach to database engineering. [11] 95–144

[3] Engels, G., Erwig, M.: ClassSheets: automatic generation of spreadsheet applica- tions from object-oriented specifications. In: Proc. of the 20th IEEE/ACM Int. Conf. on Aut. Sof. Eng., ACM (2005) 124–133

[4] Bals, J.C., Christ, F., Engels, G., Erwig, M.: ClassSheets - model-based, object-oriented design of spreadsheet applications. In: TOOLS Europe Conference (TOOLS 2007), Zürich (Swiss). Volume 6., Journal of Object Technology (October 2007) 383–398

[5] Melton, J.: Database language sql. In Bernus, P., Mertins, K., Schmidt, G., eds.: Handbook on Architectures of Information Systems. International Handbooks on Information Systems. Springer Berlin Heidelberg (1998) 103–128

[6] Google: The Google query function (version 0.7). https://developers.google.com/chart/interactive/docs/querylanguage (2014) [Accessed on February 2014].

[7] Schmidt, D.C.: Guest editor's introduction: Model-driven engineering. Computer 39(2) (February 2006) 25–31

[8] Bézivin, J.: Model driven engineering: An emerging technical space. [11] 36–64

[9] Ireson-Paine, J.: Model master: an object-oriented spreadsheet front-end. Computer-Aided Learning using Technology in Economies and Business Education (1997)

[10] Abraham, R., Erwig, M., Kollmansberger, S., Seifert, E.: Visual specifications of correct spreadsheets. In: Proceedings of the 2005 IEEE Symposium on Visual Languages and Human-Centric Computing. VLHCC '05, Washington, DC, USA, IEEE Computer Society (2005) 189–196

[11] Lämmel, R., Saraiva, J., Visser, J., eds.: Generative and Transformational Techniques in Software Engineering, International Summer School, Braga, Portugal, July 4-8, 2005. Revised Papers. In Lämmel, R., Saraiva, J., Visser, J., eds.: GTTSE 2005. Volume 4143 of Lecture Notes in Computer Science., Springer (2006)

[12] Kankuzi, B., Sajaniemi, J.: An empirical study of spreadsheet authors' mental models in explaining and debugging tasks. In: Proceedings of the 2013 IEEE Symposium on Visual Languages and Human-Centric Computing. VLHCC '13, Washington, DC, USA, IEEE Computer Society (2013) 15-18

[13] Bonnie A. Nardi. A Small Matter of Programming: Perspectives on End User Computing. MIT Press, Cambridge, MA, USA, 1993. ISBN 0262140535.

[14] Fisher, M., & Rothermel, G. The EUSES spreadsheet corpus: a shared resource for supporting experimentation with spreadsheet dependability mechanisms. ACM SIGSOFT Software Engineering Notes, 30(4), 1-5.